# Very Large Scale Integration of Josephson-Junction-Based Superconductor Random Access Memories

Vasili K. Semenov, Yuri A. Polyakov, and Sergey K. Tolpygo, *Senior Member, IEEE*

*Abstract*—Arrays of Vortex Transitional (VT) memory cells with functional density up to 1 Mbit/cm$^2$ have been designed, fabricated, and successfully demonstrated. This progress is due to recent advances in design optimization and in superconductor electronics fabrication achieved at MIT Lincoln Laboratory.

As a starting point, we developed a demo array of VT cells for the 100-µA/µm$^2$ MIT LL fabrication process SFQ5ee with 8 niobium layers. The studied two-junction memory cell with a two-junction nondestructive readout occupied 168 µm$^2$, resulting in an over 0.5 Mbit/cm$^2$ functional density. Then, we reduced the cell area down to 99 µm$^2$ (corresponding to over 0.9 Mbit/cm$^2$ functional density) by utilizing self-shunted Josephson Junctions (JJs) with critical current density, $J_c$ of 600 µA/µm$^2$ and eliminating shunt resistors. The fabricated high-$J_c$ memory cells were fully operational and possessed wide Read/Write current margins, quite close to the theoretically predicted values. We discuss approaches to further increasing the integration scale of superconductor memory and logic circuits: a) miniaturization of superconducting transformers by using soft magnetic materials; b) reduction of JJ area by using planar high-$J_c$ junctions similar to variable thickness bridges.

## I. Introduction

DECADES ago, development of superconducting latching logic circuits was considered as a more challenging task than development of Josephson junction memories utilizing the natural ability of superconductors to store data as persistent currents [1]. Since then, there has been a spectacular advance of superconductor digital electronics based on novel logic families [2], [3], [4]. Unfortunately, it has not been complemented by a similar revolution in JJ-based random access memories (RAM). The development of superconductor RAMs slowed down after the demonstration of a Vortex Transitional (VT) memory cell [5], [6] and VT-based RAM circuits in 1990s [7], [8]. The 4-Kbit level of RAM integration achieved at that time was in line with available fabrication technologies.

The reported performance of superconductor RAM also matched the expectations set by the speed of light. For instance, a 380-ps RAM access time reported in [7] matched the propagation delay in the superconducting transmission lines used for the cell selection and indicated negligible contributions of JJ switching time to the memory operation. Indeed, the speed of light in superconducting passive transmission lines (PTLs) is about $c/3 \sim 100$ µm/ps. Therefore, the minimum time required to address a memory cell in the middle of a 1-cm chip from a control circuitry at the chip edge and fetch the data is ~ 0.3 ns. This is equivalent to about 3 GHz effective clock rate of RAM.

The latter figure appears to be in a shocking mismatch with extremely high clock frequencies, tens of GHz, reported for complex RSFQ digital circuits, and reaching up to 750 GHz for the simplest ones [9]. Deep pipelining and other memory architectural solutions have been suggested to mitigate the propagation delays between logic and memory circuits [10].

In addition to the VT-based RAMs, it is appropriate to mention some other, less successful, solutions [11], [12], [13], review [14], and declared "intentions" to develop a superconductor RAM [15]. We will not discuss here many existing original proposals of single memory cells, which are either not ready yet for integration into RAM, e.g., [16], or such an integration appears to be very complicated to the authors, e.g., [17, 18], [19], etc.

The first [6] and the last [20] of the demonstrated VT cells occupied areas of 49 µm x 49 µm and 25 µm x 25 µm, respectively. The smallest VT cell was fabricated using e-beam lithography and had area of 9.5 µm x 12 µm [21]. However, this was only a standalone cell and no further development was reported.

In order to develop large-scale integration of superconductive RAMs, we started with the original VT memory cell and performed its miniaturization and optimization using the modern design tools for the now-available advanced fabrication technologies. These results are presented in Sec. II. In the following sections III and IV, we discuss approaches to further miniaturization of superconductor memory cells in order to achieve very large scale integration (VLSI) of superconductor memory required, e.g., for applications of superconducting processors in high performance computing. We mainly address

---

This research is based upon work supported by the Office of the Director of National Intelligence, Intelligence Advanced Research Projects Activity, via Air Force Contract FA872105C0002. *(Corresponding author: Vasili Semenov.)*

V. K. Semenov and Yu. A. Polyakov are with the Department of Physics and Astronomy, Stony Brook University, Stony Brook, NY 11794-3800, USA (e-mail: Vasili.Semenov@StonyBrook.edu).

S. K. Tolpygo is with the Lincoln Laboratory, Massachusetts Institute of Technology, Lexington, MA 02421, USA (e-mail: Sergey.Tolpygo@ll.mit.edu).

Color versions of one or more of the figures in this paper are available online at http://ieeexplore.ieee.org.

Digital Object Identifier will be inserted here upon acceptance.

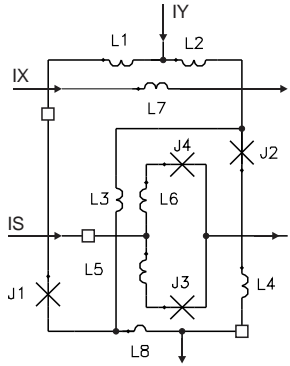

Fig. 1. Generic equivalent circuit of a VT cell. Three open squares shown in the equivalent circuit mark vias between superconductor layers. Mutual inductances between pairs of inductors (L1, L7), (L2, L7), (L3, L6), (L3, L5), and (L5, L8) were taken into account in numerical simulations. Design parameters of VT cells demonstrated in this work are given in Table I.

reduction of the area occupied by superconducting flux transformers and potential use of the planar junction technology for VLSI.

## II. CIRCUIT DESIGN AND TEST RESULTS

The circuit diagram of a generic VT memory cell is shown in Fig. 1. To better understand superconductor RAM scaling, we implemented VT cells in two fabrication technologies developed recently at MIT Lincoln Laboratory. The first, a more conventional, implementation VT1 was done in the SFQ5ee fabrication process [22], [23] and based on externally shunted Josephson junctions with $J_c = 100$ µA/µm².

The microphotograph of the fabricated VT1 memory cell is shown in Fig. 2a. We reduced the cell dimensions by a factor of two in comparison to the earlier implementations [6]-[8], [17], down to 12 µm x 14 µm, despite the evident similarity of the cell topology. This four-fold reduction in the cell area was achieved due to three factors: a) reduction of dimensional restrictions in our fabrication technology; b) reduction of shunt resistors; c) improvement of design optimization procedure. Specifically, we intentionally reduced the characteristic voltages, $I_cR_{sh}$ products, of the shunted junctions in order to reduce the area of the shunt resistors; here $I_c$ is the critical current and $R_{sh}$ is the shunt resistance. We used InductEx package [24]-[27] to accurately extract all mutual inductances, including parasitic, between submicron wires. Also, the recently developed PSCAN2 package [28] allowed us to conveniently include the mutual inductances into the numerical optimization of the cell parameters.

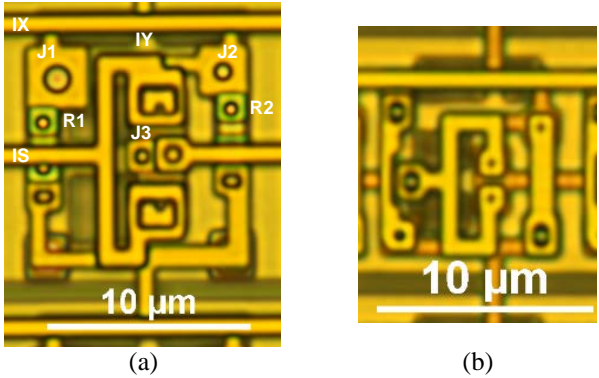

Fig. 2. Microphotographs of the fabricated memory cells. (a) Fabrication process using shunted JJs with $J_c = 100$ µA/µm²; the cell dimensions are 12 µm x 14 µm. R1 and R2 are resistor shunting junctions J1 and J2 respectively. (b) Fabrication process using self-shunted JJ with $J_c = 600$ µA/µm²; the cell dimensions are 9 µm x 12 µm. The cell layout topology in (b) corresponds to the equivalent circuit in Fig. 1.

TABLE I
DESIGN PARAMETERS OF FABRICATED MEMORY CELLS

| Design parameter / Implementation | VT1 | VT2 |
|---|---|---|
| $J_c$ (µA/µm²) | 100 | 600 |
| Memory cell size (µm) | 12 x 14 | 9 x 11 |
| Memory array pitch (µm) | 12 x 15 | 9 x 12 |
| Minimum linewidth used (µm) | 0.7 | 0.4 |
| Minimum via size / surround (µm) | 0.7 / 0.35 | 0.35 / 0.1 |
| J1 (µA) / Shunt resistor (Ω) | 193 / 2.5 | 213 / none |
| J2 (µA) /Shunt resistor (Ω) | 94 / 2.6 | 125 / none |
| J3 and J4 (µA) | 60 | 80 |
| AX[f] (µA) | 820 | 672 |
| AY[f] (µA) | 225 | 300 |
| AS[f] (µA) | 100 | 119 |
| DC (µA) | 462 | 525 |
| L1 (pH) | 7.42 | 6.10 |
| L2 (pH) | 3.17 | 1.80 |
| L3 (pH) | 7.92 | 4.33 |
| L4 (pH) | 4.75 | 2.48 |
| L5 (pH) | 3.17 | 2.77 |
| L6 (pH) | 4.75 | 2.59 |
| L8 (pH) | 1.32 | 0.98 |
| $M_{17}$ (L1, L7) (pH) | 0.95 | 1.29 |
| $M_{27}$ (L2, L7) (pH) | 0.95 | 0.58 |
| $M_{53}$ (L5, L3) (pH) | 1.32 | 0.53 |
| $M_{63}$ (L6, L3) (pH) | 1.32 | 0.90 |
| $M_{85}$ (L8, L5) (pH) | none | 0.42 |
| Write/Read transformer area (µm²)[a] | 6.7 | 5.4 |
| Area of ground plane perforation under Write/Read transformer (µm²)[b] | 18.7 | 12.4 |
| NDRO transformer wire area (µm²) | 7 | 7.2 |
| Area of ground plane perforation under NDRO transformer (µm²) | 24 | 10.5 |
| Area of ground plane moat between cells in the memory array (µm²)[d] | 5.6 | 4 |
| Memory density (Mbit/cm²) | 0.53 | 0.88 |
| Circuit density (10⁶ JJ/cm²) | 2.2 | 3.7 |

[a]Overlap area of two wires forming the transformer for Write/Read current IX.
[b]Area of an opening (moat) in the ground plane under the cell Write/Read transformer formed between inductors L1+L2 and L7.
[c]Area of a perforation in the ground plane under the nondestructive readout (NDRO) coupling transformer formed between inductors L3+L5 and L6+L8.
[d]Moat area per memory cell. These moats are used to mitigate flux trapping.
[f]AX, AY and AS are amplitudes of IX, IY and IS currents.

Design parameters of the VT1 memory cell are given in Table I. In order to reduce the size of two transformers used in the VT cells, we implemented them as coupled (overlapping) microstrips on two adjacent wiring layers. The mutual coupling between the microstrip lines was increased by making perforations (openings) in the ground plane under a large part of their length; see Sec. III. The resultant values of the most important mutual inductances, $M_{ij}$ between pairs of inductors $L_i$ and $L_j$, ($L_i$, $L_j$) are given in Table I; they were taken into account in numerical optimization of the VT cells.

The designed VT1 cells were arranged into 12 x 6 arrays with 12-µm pitch in the *x*-direction (along the rows) and 15-µm pitch in the *y*-direction. A larger pitch in the *y*-direction was used to place a 1-µm wide and 45-µm long moats in the ground plane between the adjacent rows of memory cells to mitigate magnetic flux trapping. The resultant functional density of the RAM is about 0.56 Mbit/cm$^2$.

The cells in each column of the array are connected galvanically by a bit selection line marked IY in Fig. 1. The cells in each row are coupled inductively to a common word selection line marked IX in Fig. 1. Memory Readout SQUIDs J3-J4 of all cells in each row are biased in series using a common bias line marked IS in Fig. 1. The resultant 24 independent leads are connected to contact pads. All currents are returned to the common ground plane.

The IX and IY current patterns required to perform Write and Read operations on a selected VT cell were generated and meas-

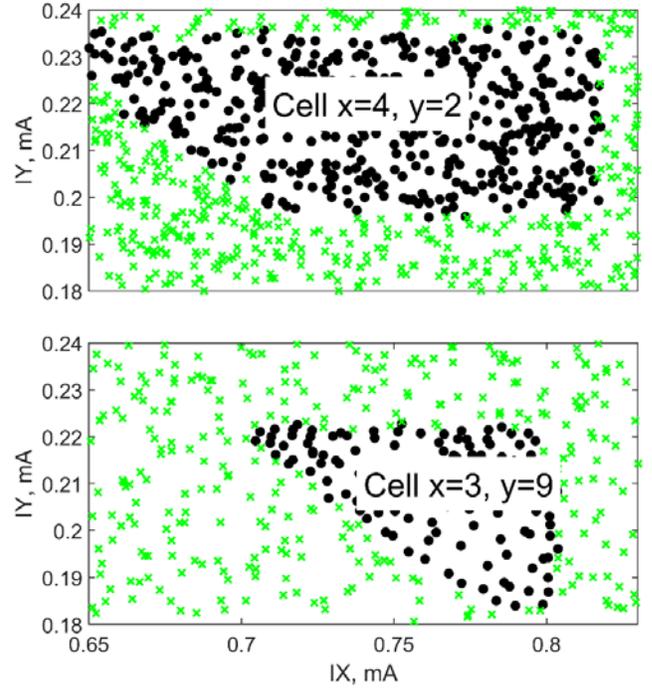

Fig. 4. Operating margins of two randomly selected VT1 cells fabricated using the 100-µA/µm$^2$ fabrication process SFQ5ee [23]. The cells are identified by their column (*x*) and row (*y*) positions in the 6x12 array. Each point in the 2D margin plots corresponds to a pair of current amplitudes |IX| and |IY| used to perform Write "0", both currents are positive, and Write "1", both currents are negative, operations. Successful operations are shown by black dots and failed operations are shown by green crosses.

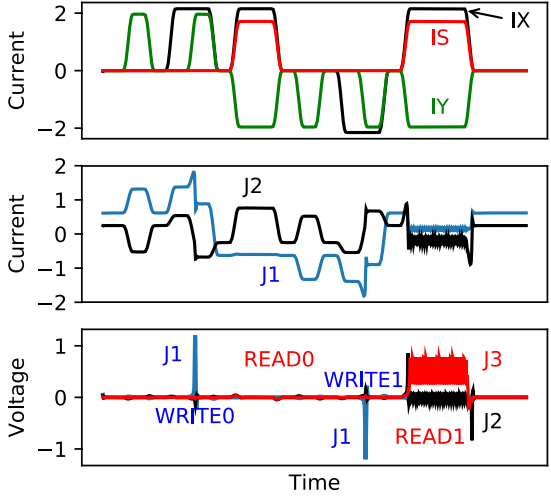

Fig. 3. Numerically simulated operation of a VT cell in Fig. 1. Three external currents shown in the upper plot initiate cell operations. Write "0" and Write "1" operations are activated by applying, respectively, positive IX (black trace) and IY (green trace) currents and negative currents IX and IY. Read operations are activated by negative IY and positive IS (red trace) currents. Superconducting currents flowing through the Josephson junctions J1 and J2 are shown in the middle panel, and voltage drops on the junctions are shown in the lower panel. The output voltage across the readout SQUID junction J3 shows Josephson oscillations in the Read "1" state.

All the traces are shown in normalized units. In order to present the test patterns in the same plot, currents IX and IS are individually rescaled, and a constant DC offset of the IX current is not shown. A substantial fraction of cell testing time is devoted to half selections when we check that applying only one and anyone current does not activate any operation. The amplitudes of IX and IY currents were varied to determine boundaries (margins) of the region of correct operation of each memory cell as shown in Fig. 4.

ured by an automated, multichannel data acquisition setup, Octopus [31] that also measured readout voltage on the IS leads. The test patterns used are functionally similar to the patterns discussed in [6]; they are shown in Fig. 3.

These 72-bit memory arrays were fabricated and tested to determine the operating margins of the individual cells. All measured cells were fully operational but demonstrated relatively narrow margins of operation shown in Fig. 4 for a few randomly selected cells. We omit further details on VT1-based memory arrays in favor of the second, more advanced implementation, VT2.

VT2 implementation is based on self-shunted JJs with 600 µA/µm$^2$ critical current density [29], [30]. Due to elimination of external shunt resistors (see, right photo in Fig. 2) and better design optimization, the cell dimensions have been reduced to 9 µm x 11 µm, giving about 41% reduction in the cell area. The cells were organized into 12 x 6 array using *x* and *y* pitches of, respectively, 9 µm and 12 µm, as shown in Fig. 5.

Fig. 6 shows test results for three randomly selected memory cells. Black circles in the Fig. 6 mark amplitudes of the IX and IY currents at successful operation cycles of the tested cells, while green crosses mark incorrect operation. The measured margins are similar to the ones projected in [5], demonstrating that performance of the fabricated circuits is in a good agreement with the theoretical model. We attribute wider margins of the VT2 cells in comparison to the ones observed on VT1 cells to a better layout optimization and parameter extraction, mainly to a better account for the mutual inductances between all the cell inductors, including parasitics – the experience gained as a result of our work on the VT1 cells.

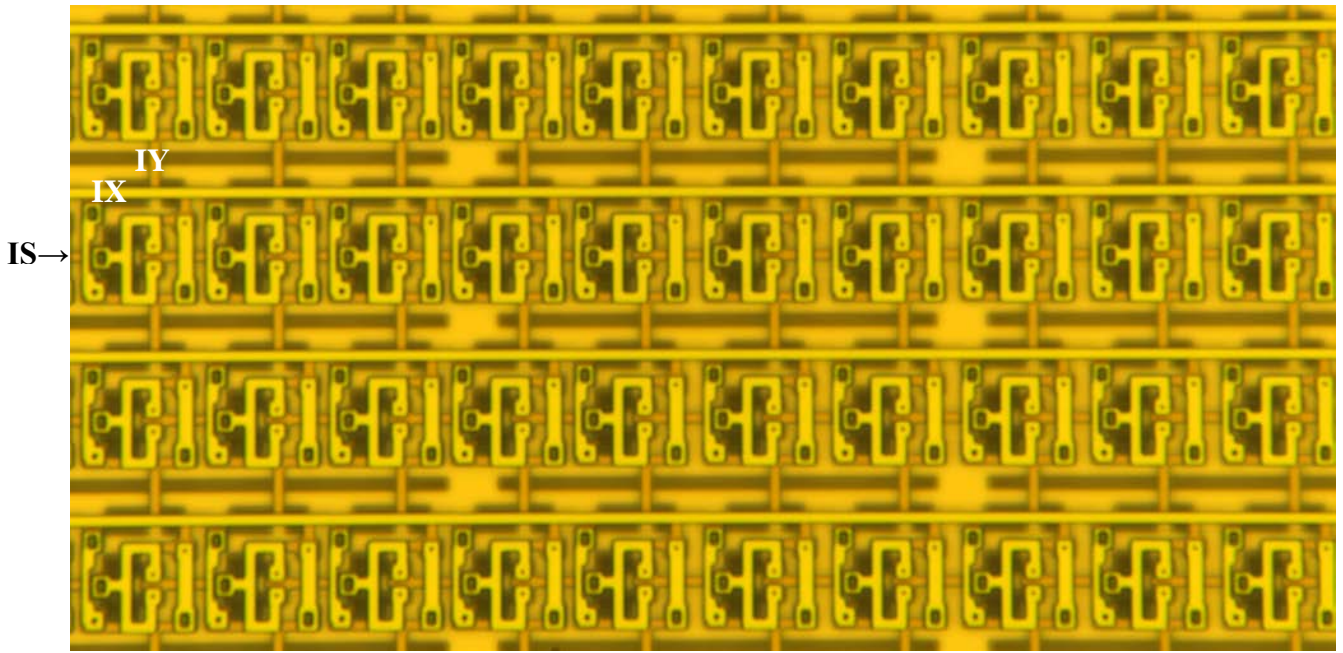

Fig. 5. A ten by four fragment of the 12 x 6 memory cell array fabricated using 600 µA/µm² technology. The word selection and bit selection lines IX and IY are marked in the picture along with the readout line IS. The chip was rotated about 90° for electrical testing, giving a difference in cell labeling in Fig. 6.

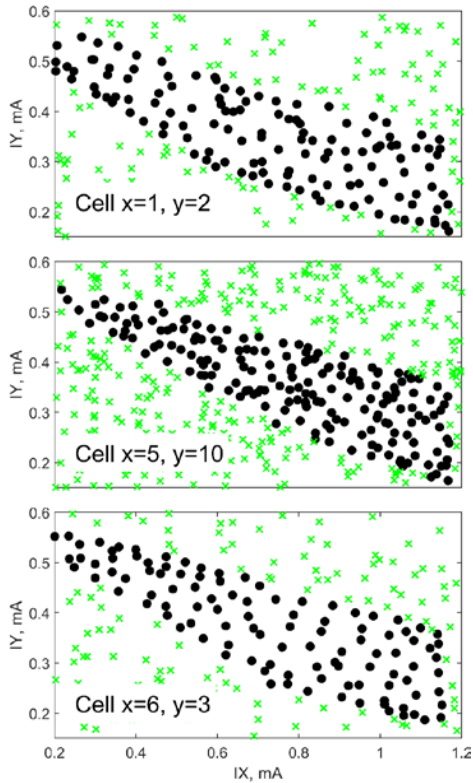

Fig. 6. Operating margins of three randomly selected VT2 cells fabricated using the 600-µA/µm² fabrication process with self-shunted junctions [29], [30]. The cells are identified by their row ($y$) and column ($x$) positions in the array consisting of 6 columns and 12 rows. Each point in the 2D margin plots corresponds to a pair of current amplitudes |IX| and |IY| used to perform Write "0", both currents are positive, and Write "1", both currents are negative, operations. Regions occupied by blue dots correspond to correct Write/Read operations.

We also performed a random pattern test on the whole array. In this test a random pattern of the size 6 x 12 was written into the array and then was read out and compared. The test was run for 100 times and showed zero errors.

To demonstrate low interference between the cells, we ran a more comprehensive, although time consuming, checkerboard test, which is used in semiconductor memory testing to detect bridging faults. The test consists of: a) writing a checkerboard pattern of "1"s and "0"s in the whole array; b) reading the content of all cells; c) writing a complementary checkerboard pattern consisting of "0"s and "1"s; d) reading the content of all cells. This test failed when applied to the whole array. However, it passed through when two particular cells were excluded from the test. The extracted cumulative margins for the checkerboard test are shown in Fig. 7. As expected, they are significantly narrower than the discussed earlier margins of individual cells.

The achieved functional density of our JJ RAM is about one million cells per cm², and the circuit density is about 4 million JJs per cm². We believe that this functional density exceeds the density of any state-of-the-art superconducting circuits reported up to date. However, the achieved density is still a few orders of magnitude below the density of SRAM and DRAM in CMOS. In the following sections, we discuss our proposals for a) doubling the functional density by incremental improvements of the existing fabrication technology; and b) more radical but still feasible technology developments required for dramatic improvements in functional density of superconductor electronics.

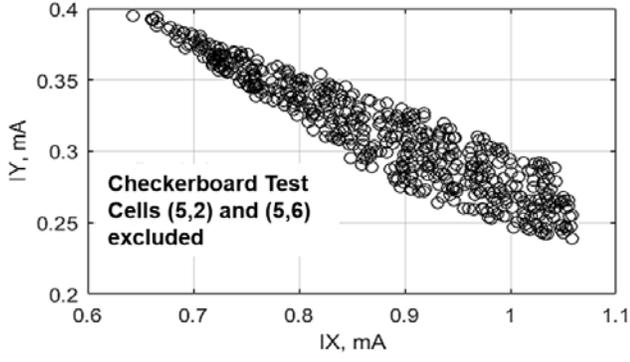

Fig. 7. Results of the checkerboard test algorithm. Cumulative margins of 70 cells in the 12 x 6 array are shown. Cells (5,2) and (5,6) failed and were excluded from the test.

## III. MINIATURIZATION OF TRANSFORMERS

### A. Superconducting Flux Transformers

About 25% of the Josephson junction-based memory cell area is occupied by two flux transformers; see Table I. The simplest superconducting transformers usually utilize a mutual inductance, $M = k(L_1 L_2)^{1/2}$ between two parallel stripline or microstrip inductors with self-inductances $L_1$ and $L_2$ located either in the same plane or stacked vertically. Spacing between the inductors and vertical separation determine the coupling coefficient, $k$. Self-inductance, $L$ and mutual inductance, $M$ of various combinations of microstrip (over one ground plane) and stripline (sandwiched between two ground planes) inductors on various layers of the MIT LL fabrication process was measured in [32] and modeled using 3D inductance extractors [24].

Since two inductors in the same plane occupy at least twice as much area as vertically stacked inductors, we used vertically stacked microstrip inductors formed by layers M5 and M6 above M4 ground plane in our memory design. For the layer stack description, Nb and interlayer dielectric thicknesses, and other parameters, see [23].

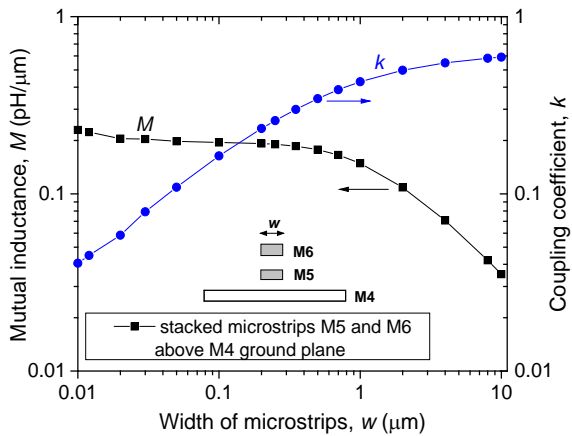

Fig. 8. Mutual inductance, $M$ per unit length (left axis) and the coupling coefficient $k = M/(L_1 L_2)^{1/2}$ (right axis) between the vertically stacked M5 and M6 inductors with the common ground plane M4. The thickness of Nb layers M4 and M6 is 200 nm and of M5 is 135 nm. The interlayer dielectric thickness between M4 and M5 is 200 nm and between M5 and M6 is 280 nm.

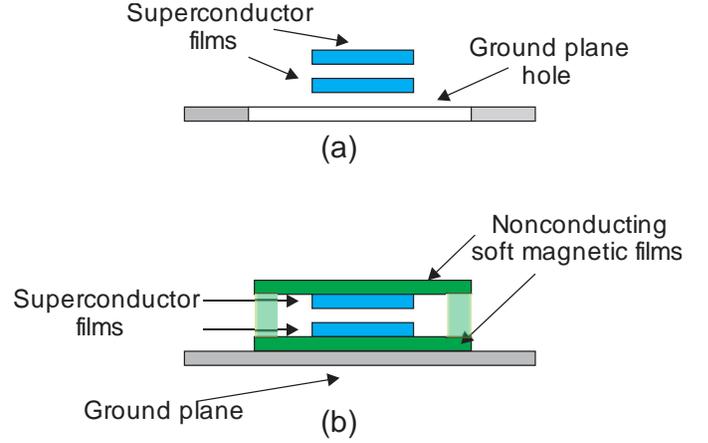

Fig. 9. (a) Cross sections of a conventional flux transformer used in our design of VT memory cells; local increase of self- and mutual inductance is achieved due to a ground plane hole. (b) A proposed transformers utilizing soft magnetic non-conducting films to increase coupling between the superconductor films. Adding vertical magnetic vias, shown in light green color, magnetically connecting the films and forming a closed magnetic core, would dramatically increase the overall transformer performance.

The mutual inductance between the stacked M5 and M6 microstrip inductors is shown in Fig. 8. The mutual inductance per unit length increases with the inductors linewidth $w$ decreasing, but saturates and becomes practically independent of $w$ below about 0.7 μm. At the same time, the self-inductance per unit length steadily increases with $w$ decreasing due to increasing of both the magnetic (geometrical) inductance and especially kinetic inductance of the film. However, the coupling coefficient $k$ strongly decreases with decreasing $w$ because more and more energy is stored as a kinetic energy of the supercurrent while magnetic flux coupling diminishes with diminishing geometrical size of the inductors.

If we want to keep IX and IY currents below 1 mA, the mutual inductance of the transformers should be about 2 pH. According to Fig. 8, this requires transformer length to be about 10 μm, independently of the width of inductors if $w < 0.5$ μm. This scale sets the dimensions of the memory cell.

It may appear that some reduction in the area of the transformers could be achieved by meandering the inductors. However, this is not possible because meander's turns need to be spaced by some nonzero amount, $s$. Therefore, the area occupied by a meander $[lw + ls(n-1) - sw(n-1) - s^2(n-1)^2/n]$ is always larger than the area $lw$ of a straight-line inductor with the same inductance, where $l$ is the straight-line inductor length and $n$ is the number of turns in the meander.

In order to increase the mutual coupling, both flux transformers in the VT cells were implemented as stacked superconductor strips crossing ground plane perforations (holes), as shown schematically in Fig. 9(a). These perforations could be seen as dark rectangles in the upper and left parts in Fig. 2 and as dark rectangles under lines IX and IS in Fig. 5.

A ground plane perforation under a strip increases its inductance because it lengthens the ground current path and also induces a circulating superconducting current around the perforation. The self-inductance in this case can be estimated as a sum of the strip inductance and the inductance of the ground plane perforation. The latter is very well known from the design of washer-type SQUIDs [37] and, for a sufficiently large ground plane, can be estimated as $\gamma\mu_0 a$, where $a$ is the size of the perforation and $\gamma$ is the perforation shape-dependent factor, e.g., $\gamma = 1.25$ for a square washer; see [26] for detailed numerical simulations for a variety of shapes. For strips with $w > (2\lambda + d)/\gamma$, where $\lambda$ is the magnetic field penetration depth and $d$ is the distance to the ground plane, the strip magnetic inductance is mainly determined by the size of the perforation.

Similarly, the mutual inductance between two microstrip inductors over the perforation is defined mostly by dimensions of the perforation rather than by the shapes of the inductors, while the coupling coefficient $k$ remains relatively low. Hence, the available design techniques do not allow for a significant downscaling of transformer dimensions.

*B. Transformers with Thin Magnetic Films*

The use of soft magnetic materials is a well-known remedy for a drastic reduction of transformer dimensions in electronics. However, it is also known that magnetic impurities can spoil or even destroy superconductivity. This is why magnetic materials have been avoided in superconductor electronics. Fortunately the prejudice against magnetic materials has been reduced due to very promising investigations of Josephson junctions with barriers made of magnetic metals; see [38], [39] for a review.

We suggest to use thin-film soft magnetic materials to reduce transformer dimensions, as illustrated in Fig. 9b. Magnetic films are located below the lower and above the upper transformer strips. Numerical simulations of the transformer using the latest InductEx package showed that a relative permeability, $\mu$ of magnetic strips of 10, or even 5, is sufficient to significantly enhance mutual inductance of the transformer in Fig. 8b. In simulations, the widths of magnetic films in Fig. 9b were taken to be equal to width of the perforation in Fig. 9a. With optional magnetic vias connecting magnetic films to form a continuous magnetic path, the size of the transformer decreases proportionally to roughly $1/\mu$. Optimization of the suggested magnetic-core transformer is still in progress and requires more details on magnetic properties of the films; the results will be reported later.

To operate at very high frequencies expected from superconductor electronics, magnetic films in the suggested transformers need to be nonconductive and possess a low imaginary part $\mu''$ of complex magnetic permeability $\mu = \mu' + i\mu''$ in order to have low RF losses in the transformer. At the same time, the real part $\mu'$ should be sufficiently high at the frequencies of operation in order to provide for a desired reduction in the transformer size. Preferably, it should also weekly depend on frequency.

There are many advanced materials, e.g., conventional ferrites, spinels, hexaferrites, etc., capable of working at up to ~ 100 GHz; see [40] for a review. Development of thin-film magnetic materials for RF applications in microelectronic integrated circuits is a very active area of industrial and academic research and subject of numerous publications, which simply could not be listed here; see [43],[44] and references therein as an example. Despite the wide use of various magnetic materials in conventional electronics, the authors have no doubts that integration of magnetic dielectrics into superconductor electronics fabrication processes is a very challenging undertaking.

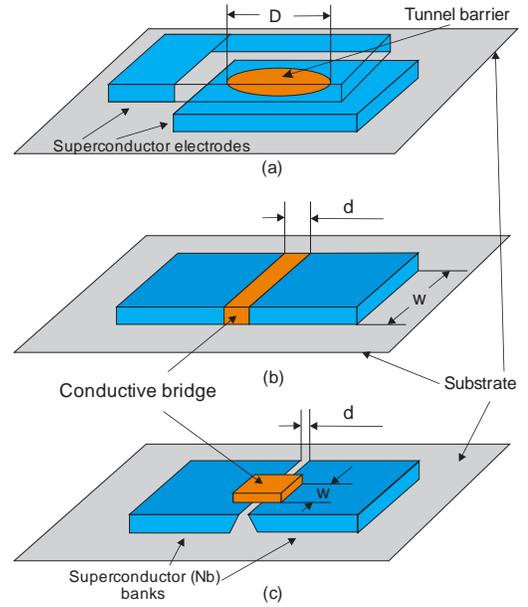

Fig. 10. A sketches of sandwich-type tunnel JJs (a), a constant-thickness (b) and variable-thickness (c) bridges consisting of a conductive material connecting two superconducting banks. For the required geometrical dimensions of Josephson bridges, see text and [34], [35].

We should note that other areas of superconductor electronics, namely various ac-biased logic circuits, would significantly benefit from the development of compact transformers. For instance, all Reciprocal Quantum Logic (RQL) [4], Quantum Flux Parametron (QFP) [40], and Adiabatic QFP (AQFP) [41] digital circuits use multiphase ac biasing schemes and flux transformers which significantly limit the achievable scale of integration.

## IV. MINIATURIZATION OF JOSEPHSON JUNCTIONS

The smallest Josephson junctions used in our VT2 cell are about 400 nm in diameter, and all four JJs occupy less than 4% of the memory cell area. Therefore, further miniaturization of the JJs does not seem to provide significant benefits to the memory density. However, the presently utilized Josephson tunnel junctions have two significant drawbacks which may complicate miniaturization of other components and integration with magnetic dielectrics. First of all, $AlO_x$ tunnel barriers used degrade above ~ 200 °C. This makes it difficult to apply modern miniaturization technologies to superconductor electronics. For instance, the widely used in CMOS industry damascene processing requires filling of high aspect ratio vias and narrow trenches by chemical vapor deposition of superconducting materials, which usually requires processing temperatures above 300 °C. Secondly, miniaturization of sandwich-type JJs shown in Fig. 10a below ~ 300 nm is difficult because it requires barriers with very high critical current density and also because JJs need a top via. Therefore, it is possible that, with progress in fabrication technology, planar bridge-type Josephson devices reviewed in [34],[35] will become competitive with sandwich-type JJs.

Among the bridge-type Josephson devices, a Variable Thickness Bridge (VTB) schematically shown in Fig. 10c looks more attractive for applications and from theoretical perspectives [34], [35] than a constant-thickness bridge (CTB) shown in Fig. 10b. Nevertheless, in some rare cases, CTBs were the only available type of devices, despite their impractical appearance; see, e.g., [47].

Basically, all theories tell that the bridge length, $d$ should be smaller than the coherence length in the bridge material, $\xi$ in order to minimize the dependence of the bridge critical current $I_c$ on $d$, which in this regime changes as $1/d$; and the shorter the bridge the better. The material-depend coherent length in thin films is typically ~ 50 nm. This sets the scale of the required bridge length.

Patterning at these dimensions is quite comfortable for the modern tools used in CMOS industry. However, if the required spread (one standard deviation, $\sigma$) of $I_c$ is, say, 2%, the $1\sigma$ spread in $d$ should be less than 1 nm, which is beyond the current capabilities even of the most expensive CMOS technologies.

The other unpleasant limitation is related to the required range of bridge resistances, $R_n$ which should be from ~ 2 Ω to ~ 8 Ω. It comes from the desired $I_c R_n$ product of ~ 0.8 mV and the typical range of $I_c$ used in digital circuits, from ~ 0.1 mA to ~ 0.4 mA. This range of $I_c$ could only be covered by changing the bridge width, $w$ since the bridge length $d$ is fixed. For the bridge material with sheet resistance, $R_s$ of 10 Ω per square, the required width is from ~ 40 nm to ~ 250 nm. If the width is defined with the same $1\sigma$ spread as the one required for $d$, the resultant spread in $I_c$ would be $\sqrt{2}\sigma$ ~ 2.8%.

These well-known estimates were given again to emphasize that implementation of the bridge-type planar technologies for superconducting VLSI circuits would require fabrication technologies as sophisticated and expensive as used in the very advanced CMOS nodes. About 35 years ago bridges and other weak links lost their competition with sandwich-like tunnel structures mainly because of the above requirements. It would be interesting to see if advantages of the planar devices for VLSI could be realized when a 50-nm patterning scale becomes standard in superconductor electronics. We note, however, that such dimensions are quite usual at a low scale of integration where such techniques as e-beam lithography and focused ion beam etching are practical, and junction parameter spreads are not important. For instance, very small Josephson junctions are mandatory components of nano-SQUIDs; see, for example, [48].

V. DISCUSSION

One of the goals of this presentation was to demonstrate advantages of the recently developed fabrication technologies [22], [23], [29], [30] for increasing the integration scale of superconductive circuits. The circuit selection was driven by the lack of JJ memories and RAMs in particular, as mentioned in Sec. I. We thought that by designing a memory circuit we could find reasons for a misfortune of conventional JJ-based memories in superconductor electronics and justify current efforts in developing various magnetic junction based cryogenic memories; see, e.g., [36], [45], and references therein. No reasons have been found. All miniaturized memory cells were fully operational and set new records for the functional density and for Josephson junction density, respectively of about 1 Mbit/cm$^2$ and $4\times10^6$ JJ/cm$^2$.

The other goal was to assess if superconductor electronics can reach integration scales and functionalities comparable to the modern CMOS circuits, a question posed in [46]. One of us (SKT) provided a negative answer in [46], based on a very general argument that superconductor electronics manipulating with circulating currents, i.e., encoding information by fluxons, cannot be miniaturized to the scale of CMOS electronics encoding information by static charge, because a localized charge would always need less space than a moving charge, especially charge moving in a loop, i.e., the circulating current. However, how far the scaling of superconductor electronics could go and at what integration scale superconducting circuits would outperform CMOS circuits remain to be answered.

By investigating the compact memory cells, we confirmed that miniaturization by linewidth reduction is the best natural remedy against parasitic flux trapping. Not only because Abrikosov vortices cannot be trapped in narrow superconducting traces, but also because effects of a magnetic flux trapped in any of the flux-trapping moats (seen as long dark horizontal rectangles in Fig. 5) become distributed between multiple cells (between eight cells in our RAM array), minimizing impact on any particular cell.

It was commonly accepted that the spacing between superconductor wires should be kept sufficiently wide to avoid their parasitic coupling. Indeed, two stripline or microstrip inductors located in the same plane couple mainly through fringing fields between them, and the $k$ nearly exponentially decreases with increasing the spacing between the inductor edges, a so-called edge coupling; see Fig. 3 in [33]. The maximum coupling $k$ ~ 0.3 is reached at zero spacing between the inductors; and $k$ weakly depends on the width of signal strips. We found that the coupling factors between wires of miniature cells remain quite low even if the wires are placed above a perforated ground plane and the wires are vertically stacked. This means that the minimum spacing is really set by the fabrication process rather than by the design.

Of course, numerous, but relatively low, parasitic couplings should be taken into account. The used tandem of InductEx and PSCAN2 perfectly executed this task. The miniaturization procedure was quite labor intensive in comparison with the past practice. This is because the stretching or narrowing of one wire affects values of all other self- and mutual inductances, i.e., the effective inductances are mutually dependent. Therefore, a known recommendation, e.g., in PSCAN2, to adjust just one inductance during circuit optimization cannot be implemented. Instead, it was necessary to run numerous optimization cycles and make a tradeoff between PSCAN2 recommendations and the real set of inductive parameters extracted by InductEx after each optimization cycle.

We hope that the proposed transformer miniaturization could be realized if technology integrating soft magnetic dielectrics with high permeability and superconductor electronics will be developed. Miniaturization of Josephson junctions is a challenging task and without the guaranteed success. Many bridge techniques have been tested with disappointing results for large-scale applications. However, bridge junctions firmly keep niche positions due to a set of unique advantages. First of all, the discussed earlier small dimensions. Secondly, being planar devices, they are better suited for vertical integrations with several stacked Josephson junction layers. Bridges could be useful even if they are formed as a narrowing in a superconductor film. For example, superconductive constrictions are used as heaters locally destroying superconductivity of thin films [16].

Our interest in Josephson junction memories was inspired in part by a recent burst of interest to programmable magnetic π-junctions and their use in "magnetic" junction memories; see, for example, MJRAM [36] and references therein. It could be noted that the equivalent circuit of a MJRAM cell in [36] is somehow similar to the schematics of a VT cell shown in Fig. 1a. In particular, the MJRAM cell explicitly contains two transformers (see Fig. 1b in [36] that shows one extra transformer hidden in Fig. 1a in [36]) and two conventional JJs. It is easy to conclude then that the area occupied by these conventional components of the MJRAM cell should be about 2/3 of the area occupied by the VT cell, if both cells are fabricated using similar technologies.

Hence, even if the magnetic JJ and its wiring does not occupy any space, the maximum expected reduction in area of the MJRAM cell with respect to the VT cell is only 30%. We believe that this would not be sufficient to justify implementation of MJRAMs, given all the technological complexities with fabrication of highly uniform programmable MJJs and high magnetic fields required for their initialization and control. Note also that these fields could be incompatible with superconductor logic circuits because of high sensitivity of the latter to magnetic field capable to induce undesirable Abrikosov or Josephson vortices. Nevertheless, we recognize that initial demonstrations may not necessarily illustrate all advantages of a new technology.

We did not attempt to implement existing or develop new drivers and decoders for JJ RAM because of a small scale of our effort. We are counting on the known solutions such as presented in [8] and potentially on a larger effort in the future.

Finally, let us note that memory cells could be also made of logic cells and, whereas transformers are highly desirable, they are not mandatory components of JJ-based memory. The key function of the transformers is to "calculate" AND functions of half-selection input data. For instance, in the VT cell, the input data are presented by DC currents IX and IY. In general, these data could arrive within a large, say, 200 ps time interval, which is too long for the RSFQ-like logic circuitry. However, S. Rylov in [49] presented an SFQ AND gate with a predefined range of time delays between input SFQ pulses, which could be utilized in a logic-cell-based JJ memories.

## VI. CONCLUSION

We have demonstrated JJ-based memory cells and RAM arrays with record functional density of about 1 Mbit/cm$^2$ and record density of Josephson junctions of $4\times10^6$ JJ/cm$^2$. We searched for new solutions to end the existing skepticism about very large scale integration of superconductor RAMs and other digital circuits. We illustrated that advancing patterning technique to 50 nm along with thin-film transformers using soft magnetic dielectrics would be sufficient to increase the functional density of RAM to 10 Mbit or even to 20 Mbit per square centimeter. Compact transformers with soft magnetic materials operating up to tens of GHz would also be highly beneficial to superconductor logic circuits such, e.g., QFP, AQFP, RQL, and others, which need transformers in their multi-phase ac-biasing networks, in signal invertors, and in logic gates.

## ACKNOWLEDGMENT

We thank Coenrad Fourie and Kyle Jackman for suggesting examples for numerical simulation of circuits with soft magnetic materials using InductEx software. We thank Pavel Shevchenko and Alex Kirichenko for the assistance with PSCAN2 software. We thank Anatoly Borodin for his assistance in measurements. We are grateful to Vladimir Bolkhovsky and Scott Zarr for their part in high-$J_c$ process development and running the process used in this work. Finally, we are thankful to Leonard Johnson, Eric Dauler, Gerald Gibson, Scott Holmes, and Marc Manheimer for their interest in and support of this work. This work was supported in part by the IARPA C3 Program. The views and conclusions contained herein are those of the authors and should not be interpreted as necessarily representing the official policies or endorsements of the U.S. Government.

## REFERENCES

[1] H. H. Zappe, "Single flux quantum Josephson junction memory cell," *Appl. Phys. Letters,* vol. 25, pp. 424-426, 1974.

[2] K. K. Likharev and V. K. Semenov, "RSFQ logic/memory family: A new Josephson-junction technology for sub-terahertz-clock-frequency digital systems," *IEEE Trans. on Appl. Supercond.,* vol. 1, pp. 3-28, Mar. 1991.

[3] D. E. Kirichenko, S. Sarwana, and A. F. Kirichenko, "Zero static power dissipation biasing of RSFQ circuits," *IEEE Trans. on Appl. Supercond.,* vol. 21, pp. 776-779, Jun. 2011.

[4] [4] Q. P. Herr, A. Y. Herr, O. T. Oberg, and A. G. Ioannidis, "Ultra-low-power superconductor logic," *J Appl. Phys.,* vol. 109, p. 8, 103903, May. 2011.

[5] S. Tahara and Y. Wada, "A vortex transitional NDRO Josephson memory cell," *Jap. J. Appl. Phys.* Part 1*,* vol. 26, pp. 1463-1466, Sep. 1987.

[6] S. Tahara, I. Ishida, Y. Ajisawa, and Y. Wada, "Experimental vortex transitional nondestructive read-out Josephson memory cell," *J. Appl. Phys.,* vol. 65, pp. 851-856, Jan. 1989.

[7] S. Nagasawa, Y. Hashimoto, H. Numata, and S. Tahara, "A 380 ps, 9.5 mW Josephson 4-kbit RAM operated at a high bit yield," *IEEE Trans. on Appl. Supercond.,* vol. 5, pp. 2447-2452, Jun. 1995.

[8] S. Nagasawa, H. Hasegawa, T. Hashimoto, H. Suzuki, K. Miyahara, and Y. Enomoto, "Design of a 16 kbit superconducting latching/SFQ hybrid RAM," *Supercond. Sci. Technol.,* vol. 12, pp. 933-936, Nov. 1999.

[9] W. Chen, A. V. Rylyakov, V. Patel, J. E. Lukens, and K. K. Likharev, "Superconductor digital frequency divider operating up to 750 GHz," *Appl. Phys. Lett.,* vol. 73, pp. 2817-2819, Nov. 1998.

[10] R. Sato, Y. Hatanaka, Y. Ando, M. Tanaka, A. Fujimaki, K. Takagi, and N. Takagi, "High-speed operation of random-access-memory-embedded microprocessor with minimal instruction set architecture based on Rapid Single-Flux-Quantum Logic," *IEEE Trans. on Appl. Supercond.,* vol. 27, 1300505, Jun. 2017.

[11] S. Nagasawa, Y. Wada, M. Hidaka, H. Tsuge, I. Ishida, and S. Tahara, "570-ps 13-mv Josephson 1-kbit NDRO RAM," *IEEE J. of Solid-State Circ.,* vol. 24, pp. 1363-1371, Oct. 1989.

[12] S. V. Polonsky, A. F. Kirichenko, V. K. Semenov, and K. K. Likharev, "Rapid single flux quantum random-access memory," *IEEE Trans. on Appl. Supercond.,* vol. 5, pp. 3000-3005, Jun. 1995.

[13] A. F. Kirichenko, S. Sarwana, D. K. Brock, and M. Radpavar, "Pipelined DC-powered SFQ RAM," *IEEE Trans. on Appl. Supercond.,* vol. 11, pp. 537-540, Mar. 2001.

[14] S. Tahara and S. Nagasawa, "Large-scale integration for high-speed Josephson random-access memory," *Appl. Supercond.,* vol. 1, pp. 1879-1891, Oct-Dec. 1993.

[15] Q. P. Herr and L. Eaton, "Towards a 16 kilobit, subnanosecond Josephson RAM," *Supercond. Sci. Technol.,* vol. 12, pp. 929-932, Nov. 1999.

[16] Q. Y. Zhao, E. A. Toomey, B. A. Butters, A. N. McCaughan, A. E. Dane, S. W. Nam, and K. K. Berggren, "A compact superconducting nanowire memory element operated by nanowire cryotrons," *Supercond. Sci. Technol.,* vol. 31, 035009, Mar. 2018.

[17] Y. Braiman, N. Nair, J. Rezac, and N. Imam, "Memory cell operation based on small Josephson junctions arrays," *Supercond. Sci. Technol.,* vol. 29, 124003, Dec. 2016.

[18] Y. Braiman, B. Neschke, N. Nair, N. Imam, and R. Glowinski, "Memory states in small arrays of Josephson junctions," *Phys. Rev. E,* vol. 94, 052223, Nov. 2016.

[19] A. Murphy, D. V. Averin, and A. Bezryadin, "Nanoscale superconducting memory based on the kinetic inductance of asymmetric nanowire loops," *New J. Phys.,* vol. 19, 063015, Jun. 2017.

[20] Y. Komura, M. Tanaka, S. Nagasawa, A. Bozbey, and A. Fujimaki, "Vortex Transitional Memory Developed with Nb 4-Layer, 10-kA/cm$^2$ fabrication process, 2015 15$^{th}$ *International Superconductive Electronics Conference (ISEC)*, paper DS-O05. DOI: 10.1109/ISDEC.2015.7383489

[21] H. Numata, S. Nagasawa, and S. Tahara, "A vortex transitional memory cell for 1-Mbit/cm$^2$ density Josephson RAMs," *IEEE Trans. Appl. Supercond.,* vol. 7, pp. 2282-2287, Jun. 1997.

[22] S. K. Tolpygo, V. Bolkhovsky, T. J. Weir, L. M. Johnson, M. A. Gouker, and W. D. Oliver, "Fabrication process and properties of fully-planarized deep-submicron Nb/Al-AlO$_x$/Nb Josephson junctions for VLSI circuits," *IEEE Trans. Appl. Supercond.,* vol. 25, 1101312, Jun. 2015.

[23] S. K. Tolpygo, V. Bolkhovsky, T. J. Weir, A. Wynn, D. E. Oates, L. M. Johnson, and M. A. Gouker, "Advanced fabrication processes for superconducting very large-scale integrated circuits," *IEEE Trans. on Appl. Supercond.,* vol. 26, 1100110, Apr. 2016.

[24] C. J. Fourie, O. Wetzstein, T. Ortlepp, and J. Kunert, "Three-dimensional multi-terminal superconductive integrated circuit inductance extraction," *Supercond. Sci. Technol.,* vol. 24, 125015, Dec. 2011.

[25] C. J. Fourie, O. Wetzstein, J. Kunert, H. Toepfer, and H. G. Meyer, "Experimentally verified inductance extraction and parameter study for superconductive integrated circuit wires crossing ground plane holes," *Supercond. Sci. Technol.,* vol. 26, 015016, Jan. 2013.

[26] C. J. Fourie, A. Takahashi, and N. Yoshikawa, "Fast and accurate inductance and coupling calculation for a multi-layer Nb process," *Supercond. Sci. Technol.,* vol. 28, 035013, Mar. 2015.

[27] C. J. Fourie, "Calibration of inductance calculations to measurement data for superconductive integrated circuit processes," *IEEE Trans. on Appl. Supercond.,* vol. 23, 1301305, Jun. 2013.

[28] P. Shevchenko. (2015). *PSCAN2 Superconductor Circuit Simulator.* Available: http://www.pscan2sim.org/

[29] S. K. Tolpygo, V. Bolkhovsky, S. Zarr, T. J. Weir, A. Wynn, A. L. Day, L. M. Johnson, and M. A. Gouker, "Properties of unshunted and resistively shunted Nb/AlO$_x$/Al/Nb Josephson junctions with critical current densities from 0.1 mA/μm$^2$ to 1 mA/μm$^2$," *IEEE Trans. on Appl. Supercond.,* vol. 27, no. 4, 110815, Jun. 2017.

[30] S. K. Tolpygo, V. Bolkhovsky, D. E. Oates, R. Rastogi, S. Zarr, A. L. Day, T. J. Weir, A. Wynn, and L. M. Johnson, "Superconductor electronics fabrication process with MoN$_x$ kinetic inductors and self-shunted Josephson junctions," *IEEE Trans. on Appl. Supercond.,* vol. 28, 1100212, Jun. 2018.

[31] D. Y. Zinoviev and Y. A. Polyakov, "Octopux: An advanced automated setup for testing superconductor circuits," *IEEE Trans. Appl. Supercond.,* vol. 7, pp. 3240-3243, Jun. 1997.

[32] S. K. Tolpygo, V. Bolkhovsky, T. J. Weir, C. J. Galbraith, L. M. Johnson, M. A. Gouker, and V. K. Semenov, "Inductance of circuit structures for MIT LL superconductor electronics fabrication process with 8 niobium layers," *IEEE Trans. Appl. Supercond.,* vol. 25, 1100905, Jun. 2015.

[33] S. K. Tolpygo, V. Bolkhovsky, R. Rastogi, S. Zarr, A. L. Day, T. J. Weir, A. Wynn, and L. M. Johnson, "Developments toward a 250-nm, fully planarized fabrication process with ten superconducting layers and self-shunted Josephson junctions," *in IEEE Trans. Appl. Supercond. 2017 16th International Superconductive Electronics Conference*, 2017.

[34] K. K. Likharev, "Superconducting weak links," *Rev. Mod. Phys.,* vol. 51, pp. 101-159, 1979.

[35] A. A. Golubov, M. Y. Kupriyanov, and E. Il'ichev, "The current-phase relation in Josephson junctions," *Rev. Mod. Phys.,* vol. 76, pp. 411-469, Apr. 2004.

[36] I. M. Dayton, T. Sage, E. C. Gingrich, M. G. Loving, T. F. Ambrose, N. P. Siwak, S. Keebaugh, C. Kirby, D. L. Miller, A. Y. Herr, Q. P. Herr, and O. Naaman, "Experimental demonstration of a Josephson magnetic memory cell with a programmable pi-junction," *IEEE Magn. Lett.,* vol. 9, 3301905, 2018.

[37] J.M. Jaycox and M.B. Ketchen, "Planar coupling scheme for ultra low noise dc SQUIDs," *IEEE Trans. Mag.*, vol. MAF-17, no. 1 pp. 400-403, Jan. 1981.

[38] A.I. Buzdin, "Proximity effects in superconductor-ferromagnet heterostructures," *Rev. Mod. Phys.*, vol. 77, p. 935, 2005.

[39] V.V. Ryazanov, "Physics and application of superconducting phase inverters based on superconductor-ferromagnet-superconductor Josephson junctions," in *Fundamentals of Superconducting Nanoelectronics*: A. Sidorenko (eds), Springer-Verlag, Berlin, 2011, pp. 219-248.

[40] M. Pardavi-Horvath, "Microwave applications of soft ferrites," *J. Magnetism & Magnet. Mater.*, vol. 215-216, pp. 171-183, 2000.

[41] M. Hosoya, W. Hioe, J. Casas, "Quantum flux parametron: a single quantum flux device for Josephson supercomputer," *IEEE Trans. Appl. Supercond.*, vol. 1, No. 2, pp. 77–89, Jun. 1991.

[42] N. Takeuchi, D. Ozawa, Y. Tamanashi, and N. Yoshikawa, "An adiabatic quantum flux parametron as an ultra-low-power logic device," *Supercond. Sci. Technol.*, vol. 26, No. 3, Mar. 2013, Art. no. 035010.

[43] A. El-Chazaly, R.M. White, S. X. Wang, "Gigahertz-band integrated magnetic inductors," *IEEE Trans. Microwave Theor. Technol.*, vol. 65, no. 12, pp. 4893-4900, Dec. 2017.

[44] P. K. Amiri, B. Rejaei, Y. Zhuang, M. Vroubel, D.W. Lee, S.X. Wang, and J.N. Burghartz, "Integrated microstrip lines with Co-Ta-Zr magnetic films," *IEEE Trans. Mag.*, vol. 44, no. 11, pp. 3103-3106, Nov. 2008.

[45] I.P. Nevirkovets and O.A. Mukhanov, "Memory cell for high-density arrays based on a multiterminal superconducting-ferromagnetic device," *Phys. Rev. Appl.*, vol. 10, 034013, Sep. 2018.

[46] S.K. Tolpygo, "Superconductor digital electronics: Scalability and energy efficiency issues (Review Article)," *Low Temp. Phys.*, vol. 42, no. 5, pp. 361-379, May 2016.

[47] S. Shokhor, B. Nadgorny, M. Gurvitch, V. Semenov, Y. Polyakov, K. Likharev, S. Y. Hou, and J. M. Phillips, "All-high-Tc superconductor rapid-single-flux-quantum circuit operating at similar-to-30-K," *Appl. Phys. Lett.,* vol. 67, pp. 2869-2871, Nov. 1995.

[48] W. Wernsdorfer, "From micro- to nano-SQUIDs: applications to nanomagnetism," *Supercond. Sci. Technol.*, vol. 22, 064013, 2009.

[49] S. Rylov, "Clockless dynamic SFQ (DSFQ) AND gate with high input skew tolerance," *IEEE Trans. Appl. Supercond.* (Early Access). DOI: 10.1109/TASC.2019.2896137